# Experimental Demonstration of the Co-existence of the Spin Hall and Rashba Effects in beta-Tantalum/Ferromagnet Bilayers

**Authors:** Gary Allen[1*], Sasikanth Manipatruni[1], Dmitri E. Nikonov[1], Mark Doczy[1] and Ian A. Young[1]

**Affiliations**

[1] Components Research, Intel Corp., Hillsboro, Oregon, 97124, USA.

[*] Corresponding author. E-mail: gary.allen@intel.com

**Abstract**

We have measured the spin torques of beta-Tantalum / $Co_{20}Fe_{60}B_{20}$ bilayers versus Ta thickness at room temperature using an FMR technique. The spin Hall coefficient, $\theta_{sh}$, was calculated both from the observed change in damping coefficient of the ferromagnet with Ta thickness, and from the ratio of the symmetric and anti-symmetric components of the FMR signal. Results from these two methods yielded values for $\theta_{sh}$ of -0.090±0.005 and -0.11±0.01, respectively. We have also identified a significant out-of-plane spin torque originating from Ta, which is constant with Ta thickness. We ascribe this to an interface spin orbit coupling, or Rashba effect, due to the strength and constancy of the torque with Ta thickness. From fitting measured data to a model including interface spin orbit coupling, we have determined the spin diffusion length for beta-Tantalum to be ~2.5 nm.

**Introduction**

Complementary metal-oxide-semiconductor (CMOS) integrated circuits have been scaled over the past four decades and will continue for at least another 15 years (*1*). At the same time, research in beyond-CMOS devices (*2,3,4*) is being actively pursued. The main drivers in this research is the need for low-power logic circuits (*5*) as well as non-volatile circuits enabling normally-off instantly-on computing systems (*6*). One of the most explored options for beyond CMOS devices is spintronic (nanomagnetic) logic (*7,8,9*).

First commercial spintronic devices, magnetic RAM (*10*), operated by switching magnetization via the Oersted magnetic field of a current. More efficient switching by spin transfer torque (STT) (*11,12*) gave rise to a new generation of STT-RAM (*13*) which operates at a smaller required switching current. Spin Hall effect (SHE) was theoretically predicted (*14*) and more recently applied to switching magnetization (*15*). It is expected to switch magnetization at an even smaller current. Understanding the properties of these materials and SHE parameters in them is necessary for estimating the performance of devices and circuits comprising them.

SHE has been observed in elemental metals with high atomic number: Pt (*15*), W (*16*), and Ta (*17*), and more recently in a topological insulator $Bi_2Se_3$ (*18*). The effect originates from spin-orbit coupling in these materials and is manifested as creation of spin polarized electrons at the surface of the material (*19*). If a SHE material is in contact with a ferromagnet (FM), the spin polarized current is injected into it and is capable of spin transfer torque switching its magnetization direction. The inverse spin Hall effect (ISHE) (*19*) is the generation of charge current in such materials when spin polarized current is injected in them, e.g. from a ferromagnet. Typically SHE is probed by the method of ferromagnetic resonance (*20*), i.e. by applying a RF magnetic field to the stack of SHE and FM materials, scanning the value of a DC magnetic field, and measuring a produced DC voltage.

The important difference between these materials is that Pt has a conductivity much higher than that of a typical FM, while Ta has a conductivity comparable to that of a FM. Consequently, several spin-orbit coupling related phenomena are convolved in the case of Ta. These are: 1) precession of magnetization in the RF magnetic field and the associated change in the anisotropic magnetoresistance (*21,22*); 2) the spin Hall torque (*19*) affecting the magnetization precession; 3) injection of spin-polarized electrons from FM to the SHE material (spin pumping) (*23*) and an induced charge current by ISHE (*19*); and 4) spin Hall magnetoresistance (*24*). For Ta, four different published values of SHE coefficient (*25,17,26,27*) span a range of more than an order of magnitude. One study of the SHE dependence on Ta thickness (*28*) can be found in literature, while several publications (*29,30,31,32,33,34*) address the thickness dependence of SHE in Pt or Bi (*35*). In this paper we study Ta SHE at various thicknesses with the aim of de-convolving the above phenomena and making an accurate determination of the SHE coefficient.

We describe a model for the dynamics of magnetization to explain the experimental results, the procedure for fabrication of samples, the experimental method, and the results of the measurement. We close with the discussion and interpretation of the experimental results.

**Model of Spin-Orbital Effects**

To describe the experimental results presented in this paper, we solve the Landau-Lifshitz-Gilbert (LLG) equation for the magnetization of a FM in an oscillating magnetic field that is in contact with a normal metal (NM) exhibiting the SHE. We include in our derivation the often neglected out-of-plane torque term. The full LLG is,

$$\frac{d\hat{\mathbf{m}}}{dt} = -\gamma \hat{\mathbf{m}} \times \mathbf{H}_{eff} + \alpha \hat{\mathbf{m}} \times \frac{d\hat{\mathbf{m}}}{dt} + \gamma \beta_{\parallel} \hat{\mathbf{m}} \times [\hat{\mathbf{s}} \times \hat{\mathbf{m}}] + \gamma \beta_{\perp} \hat{\mathbf{s}} \times \hat{\mathbf{m}}, \qquad (1)$$

where *m* is the unit vector of the FM's magnetization and *s* is the unit vector of the polarization of the spin current injected into the FM from the NM. The quantities $H_{eff}$, $\gamma$ and $\alpha$ are the effective magnetic field (applied field plus demagnetization field), the gyromagnetic ratio and the damping coefficient, respectively. Lastly, the two quantities, $\beta_{//}$ and $\beta_{\perp}$, are the coefficients for the anti-damping (in-plane) and field-like (out-of-plane) spin torques, respectively, and have units of magnetic field:

$$\beta_{\parallel} \equiv \varepsilon_{\parallel} \frac{\hbar}{2e} \frac{J_s}{M_s t_m}, \quad \beta_{\perp} \equiv \varepsilon_{\perp} \frac{\hbar}{2e} \frac{J_s}{M_s t_m}. \qquad (2)$$

The coefficients $\varepsilon_{//}$ and $\varepsilon_{\perp}$ are introduced to take into account the efficiency of the spin transfer process (*36*). Spin current densities, $J_s$, are in units of amperes/m$^2$.

In the presence of an applied magnetic field, the direction of the magnetization of a FM will precess around the direction of the applied magnetic field, at an angle of precession, $\phi_p$. This typically small change in the direction of magnetization, $\Delta m$, can be calculated from Eq. 1. Since the angle of precession is small, it can be approximated as $\Delta m$, which from Eq. 1 is,

$$\phi_p \approx \Delta m = \frac{1}{\Delta(2H_0 + M_s)}\left[\left(h_y^{rf} + \beta_{\perp} s_y\right) B_0 \mathrm{L_A}(H) + \left(\beta_{\parallel} \frac{\omega}{\gamma} s_y\right) \mathrm{L_S}(H)\right], \qquad (3)$$

where $\Delta$ is the line width of FMR, $h_y^{rf}$ is the y-component of the RF magnetic field, $s_y$ is the y-component of the spin polarization unit vector, and $H_0$ is the applied field at resonance. The functions $L_S$ and $L_A$ are the symmetric and anti-symmetric Lorentzians, respectively:

$$L_a(H) = \frac{\Delta(H-H_0)}{(H-H_0)^2 + \Delta^2} \quad , \quad L_s(H) = \frac{\Delta^2}{(H-H_0)^2 + \Delta^2}. \tag{4}$$

We have assumed here that, due to the large in-plane anisotropy of the thin FM, the precession is primarily in the film plane. and the precession of the magnetization is elliptical.

**Ferromagnetic Resonance Measurements**

To measure the SHE of Ta, we use the FMR measurement technique developed by Liu et al. (*15*). In their technique, an RF charge current is sent through a FM|NM bilayer at GHz frequencies. This arrangement of layers and driving frequency has the advantage of emulating that of proposed SHE driven spintronic devices. (e.g. the top two layers of an MTJ driven at GHz frequencies).

When this current is injected into a bilayer line, the portion of the current that flows through the NM layer simultaneously creates an in-plane RF magnetic field (the Oersted field) in the FM, as well as, injecting a spin current into the FM due to the SHE. Both of these effects influence the precession of the magnetization around the direction of the applied field. The oscillation of the angle of magnetization from precession causes an oscillation of the resistance of the FM due to AMR. Since the frequency of the oscillation is the same as that of the applied RF current, frequency mixing produces a dc voltage that is proportional to both the current flow in the FM and its magneto-resistance.

A schematic of our measurement setup is given in Fig. S1, where a vector network analyzer is shown to force a microwave frequency current into a Ta/CoFeB bilayer. The resulting DC voltage across the length of the line is on the order of microvolts or less, so a lock-in amplifier is used in conjunction with an RF switch that chops the RF current at 1.37kHz.

The samples' lines form the center conductor of a coplanar waveguide, with the ground lines and probe pads made of Au. Electrical contact is made with 150um pitch GSG probes. Measurements were made at frequencies of 7, 8, 9 and 10 GHz with 17dBm±0.5dBm of RF power at the probe tips. The generated DC voltage was measured as a function of an applied in-plane magnetic field aligned 45° to the direction of the RF current. All measurements were made at room temperature.

**Measurement Results**

Using the expression for the precession angle of the magnetization in Eq. 3, the DC voltage as a function of applied field strength can be shown to equal,

$$V_{DC} = -\frac{1}{2} I_{rf} \Delta R_{AMR} \frac{\sin(2\phi_0)\cos\phi_0}{\Delta(2H_0 + M_s)} \frac{\omega}{\gamma} \left[ \sqrt{\tfrac{B_0}{H_0}}(h_{rf} + \beta_\perp)L_a(H) + \beta_\| L_s(H) \right]$$
$$\equiv V_A L_A(H) + V_S(H) L_S(H) \tag{5}$$

From Ampere's law, the Oersted field in the FM created by the NM is $h_{rf} = J_c^{NM} t_{NM}/2$.

In Fig. 1, the measured DC voltage as a function of applied field strength is shown for four different Ta thicknesses at a driving frequency of 7 GHz. The measured voltage versus applied field is fitted to the sum of a symmetric and an anti-symmetric Lorentzian, in accordance to Eq. 5. The parameters of the fit are the weights of the Lorentzians, $V_A$ and $V_S$, and the resonant applied field and line-width, $H_0$ and $\Delta$. There is also an offset present in the data that is not encompassed in Eq. 5 that we calculate only to remove it from the data. Data for the symmetric and anti-symmetric components of the DC voltage versus Ta thickness, $V_S$ and $V_A$ respectively, at a 7 GHz driving frequency are plotted in Fig. 2.

The fit values for the resonant field, $H_0$, versus the driving frequency can be fit to the Kittel equation for thin films (*20*), $\omega^2 = \gamma\sqrt{H_0(H_0 + M_s)}$, to determine the saturation magnetization, $M_s$, of the CoFeB layer. The value of $M_s$ for the CoFeB only sample, 1.92±0.01T, was found to be larger than that of the samples deposited with Ta, 1.84±0.01T averaged over six samples (Fig. S2). It is known that Ta sputtered onto CoFeB can create a magnetic dead layer (*37*), which itself will not change the magnetization (i.e. the magnetic moment per unit volume is unchanged), so we ascribe the reduction in $M_s$ to an intermediate layer at the interface in which the magnetization CoFeB is reduced but not eliminated. For comparison, we measured the 5.3nm Ta sample on a SQUID magnetometer and obtained a value for $M_s$ of 1.86±0.05T, which is in agreement with the values from FMR measurement.

The damping coefficient, α, of the CoFeB layer can be determined from the fit values for the line-width, $\alpha = \gamma\Delta/\omega$, the results of which are shown in Fig. 3A. As can be seen, α decreases with Ta thickness, reaching a minimum value at 3.3nm, and increases slightly for larger thicknesses. We attribute this thickness dependence to anti-damping in the CoFeB due to spin torque from the SHE spin current produced by Ta. As has been shown in references (*38,39,17*), the measured damping coefficient in the presence of a spin current can be attributed to the intrinsic value of the FM plus a contribution due to the damping/anti-damping from the SHE spin current, and can be shown to equal,

$$\gamma\Delta/\omega = \alpha_0 + \beta_\parallel \frac{2\sin\phi}{2H_0 + M_s}. \qquad (6)$$

The SHE coefficient of the Ta layer can be calculated from the damping data by first determining the spin current density, $J_s$, from equations (2) and (6). The charge current density in Ta, $J_c$, can be calculated from the measured data using equation (5). We found $J_c$ to be approximately equal for all samples at all applied RF frequencies used, $1.11\pm0.09\times10^{10}$ A/m$^2$. The SHE coefficient, $\theta_{sh}$, is then the ratio of $J_s$ and $J_c$, which is plotted in Fig. 3B.

A second method for calculating the SHE coefficient uses the data for the voltage contribution of the symmetric, $V_S$, and anti-symmetric, $V_A$, components of the DC voltage versus Ta thickness. The values of these quantities are calculated from curve fitting the measured DC voltage versus applied field to Eq. 5. Values for $V_S$ and $V_A$ are shown in Fig. 2 for a driving frequency of 7GHz. The statistical error (2σ) for all values of $V_S$ and $V_A$ measured at the four frequencies was always less than 0.28µV.

Assuming a perfectly spin conducting interface, the data for $V_S$ is expected to follow the relation; $V_S(t_{NM}) = V_S^\infty\left[1 - \text{sech}\left(\frac{t_{NM}}{\lambda_{sf}}\right)\right]$, with $V_S^\infty$ being the large thickness limit of $V_S$. This value is indicated by the dashed green line in Fig. 2. The dependence of $V_A$ on Ta thickness is linear, but has a positive offset. This results in $V_A$ becoming zero and changing sign at a Ta thickness of 1.7nm. This effect is also clearly seen in Fig. 1 where the anti-symmetric component of the fit to data is seen to change sign between the 1.3nm and 2.3nm sample. We ascribe the positive offset in $V_A$ with thickness to an out-of-plane spin torque originating from an interface spin orbit coupling effect (also referred to as Rashba effect), which we explained later in this letter.

The spin Hall coefficient, $\theta_{sh}$, can be calculated from $V_S^\infty$ and the slope of $V_A$ versus Ta thickness, $V_A^{slope}$, from the equation,

$$\theta_{sh} = \left(\frac{V_S^\infty}{V_A^{slope}}\right)\sqrt{1 + \frac{M_s}{H_0}} \frac{e}{\hbar} \mu_0 M_s t_{FM} . \qquad (7)$$

For the data shown in Fig. 2, and data taken at frequencies of 8, 9 and 10 GHz, we calculate a spin Hall coefficient for $\beta$-Ta of -0.11±0.01. This value is slightly higher, but in reasonable agreement with the $\theta_{sh}$ determined using the change in α due to anti-damping shown in Figure 3(b).

This value is lower than that of 0.15±0.03 reported by (*17*) for the same measurement technique. This is due to the fact that in (*17*), the value was based on only a single Ta thickness, and hence, no accounting of the offset in $V_A$ versus Ta thickness could be included. Using their expression for $\theta_{sh}$ and using the value for $V_A$ only, we calculate for a similar thickness of Ta and CoFeB, Ta(7.3)/CoFeB(4), a value of -0.13±0.03, indicating that both measurements are in agreement, if not in interpretation. Our value for $\theta_{sh}$ is in good agreement with the two other values reported in (*17*), both of which are -0.12±0.03, which were obtained from the dependence of damping on a DC bias current, and the critical current to switch a magnetic tunnel junction device.

Compared to other published data, our value is considerably less than the value of -0.19 (no errors reported) reported by (*26*), which was obtained from DC measurement. It is also an order of magnitude larger than the values of -0.02(+0.008,-0.015) reported by (*27*) and 0.0037±0.0011 reported by (*25*). It should be noted that the value reported in (*27*) is for Ta/YIG bilayer, and that reported in (*25*) is for Ta to Cu spin channel to permalloy system. It is reasonable to expect that the spin current transferred across a NM/FM interface is dependent on the materials at the interface, and could thus explain the differences in reported values of $\theta_{sh}$ for Ta. A derivation proof, from our measured data, for the existence of this interface effect is discussed in the next section.

**Explanation of the Thickness Dependence of the Spin Torques**

As shown in Eq. 5, the symmetric and anti-symmetric components of the measured DC voltage originate from anti-damping and field-like spin torques, respectively. We attribute the thickness dependence of the spin torques as arising from spin currents produced by both the SHE of the bulk β-Ta and the Rashba effect at the Ta/CoFeB interface. In particular, we identify the Rashba effect as the source of the positive offset of the anti-symmetric component seen in Fig. 2.

A similar offset in the thickness dependence of the Inverse SHE current from spin pumping measurements was observed by Hou *et al.* (*35*) in bismuth/permalloy bilayers. They were able to explained the offset as originating from a distinct interface layer with a spin Hall coefficient and spin diffusion length different from that of the bulk material. However, in the case of Ta/CoFeB bilayers measured in this study, we can explain the origin of the offset to the spin current produced at the interface by the Rashba effect.

We use a semi-classical model (*40*) which uses a Boltzmann equation to model the presence of an interface spin orbit coupling (ISOC) effect (referred to as Rashba effect) and a bulk spin orbit coupling effect (BSOC) (referred to as Spin Hall Effect). The semi-classical model comprehends a) BSOC and ISOC, b) effect of finite ($G_{mix}$) spin-mixing conductance between the FM and NM, c) effect of relative ratio of imaginary ($Im(G_{mix})$) and real ($Im(G_{mix})$) components of spin mixing conductance, d) thickness effect due to scaling of spin mixing conductance which depends on the available states in the NM ($\tilde{G}_{mix}(t) = G_{mix} 2\rho_{FM} l_{sf} \tanh(t/l_{sf})$), e) thickness effect due to the spin diffusion in the NM.

We write the interfacial torque on the FM (assuming spin current absorption at the interface as $\lambda_{sf\text{-}FM}$ approaches zero) (*40*)

$$T = \delta(z) \frac{g\mu_B j_0}{2e} \left[\tau_d \hat{M} \times (\hat{M} \times \hat{y}) + \tau_f \hat{M} \times \hat{y}\right] \tag{8}$$

where $\hat{y}$ is the direction of spin moment of the injected electrons propagating along $\hat{z}$, the interface vector for FM to NM interface (Fig. 4).

We note that the ISOC as well as BSOC can generate damping and field like torques. BSOC generates predominately (anti) damping torque. However BSOC can generate a field like torque component proportional to the imaginary part of spin mixing conductance of the NM to FM interface. ISOC generates predominantly field like torque. However, ISOC can also generate a damping torque depending on the exact dephasing mechanism inside the FM (*40*). However, it has been noted that while both ISOC and BSOC contribute to both $\tau_d$ and $\tau_f$ only, BSOC exhibits a strong dependence on the thickness of the NM.

We include both the ISOC and BSOC contributions to the damping and field like contribution to the spin torques to explain the torque dependence in a FM/NM bilayer. The damping torque in presence of ISOC and BSOC can be written as,

$$\tau_d = \tau_{dBSOC} + \tau_{dISOC} = \theta_{SHE} \frac{\left(1-e^{-t/\lambda_{sf}}\right)^2}{\left(1+e^{-2t/\lambda_{sf}}\right)} \times \left[\frac{|\tilde{G}^{\uparrow\downarrow}|^2 + \text{Re}[\tilde{G}^{\uparrow\downarrow}]\tanh^2(t/\lambda_{sf})}{|\tilde{G}^{\uparrow\downarrow}|^2 + 2\text{Re}[\tilde{G}^{\uparrow\downarrow}]\tanh^2(t/\lambda_{sf}) + \tanh^4(t/\lambda_{sf})}\right] + \tau_{dISOC} \tag{9}$$

where, we included a NM-thickness independent contribution to the damping spin torque arising from ISOC. $\tilde{G}_{mix}(t)$ is a scaled spin mixing conductance accounting for thickness induced effects. The field like torque in presence of ISOC and BSOC can be written as,

$$\tau_f = \tau_{fBSOC} + \tau_{fISOC} + \tau_{fAmpere} =$$

$$\theta_{SHE} \frac{\left(1-e^{-t/\lambda_{sf}}\right)^2}{\left(1+e^{-2t/\lambda_{sf}}\right)} \times \left[\frac{\text{Im}[\tilde{G}^{\uparrow\downarrow}]\tanh^2(t/\lambda_{sf})}{|\tilde{G}^{\uparrow\downarrow}|^2 + 2\text{Re}[\tilde{G}^{\uparrow\downarrow}]\tanh^2(t/\lambda_{sf}) + \tanh^4(t/\lambda_{sf})}\right] + \tau_{fISOC} + \frac{CV}{\rho(L/w)} t \tag{10}$$

where the third term is the ampere field due to the current. A fixed thickness independent contribution is added due to the field like contribution from interface spin orbit effect.

To deconvolve what could be happening, we first consider the possibility of a BSOC only explanation for the measured damping and field like torques as a function of HM electrode thickness. We vary the reflectivity ratio R from 1 to 0.2, where R is defined as the ratio

$$R \equiv \frac{\text{Im}(G_{mix})}{|G_{mix}|} \tag{11}$$

Fig. 4 shows the expected dependence of BSOC assuming the presence of just a BSOC and no ISOC. The damping like torque ($\tau_{dBSOC}$) from BSOC saturates to a maximum value at a thickness t>$\lambda_{sf}$. The field like torque from BSOC ($\tau_{fBSOC}$) exhibits a similar saturation near t>$\lambda_{sf}$. In an experimental measurement using FMR technique this would result in a constant offset only at t>$\lambda_{sf}$ as shown in Fig. 4. Fig. 4A shows the case R=1 implying that the spin mixing conductance is purely imaginary. The field and damping like torque from BSOC are comparable in this case both approaching the same value at t>$\lambda_{sf}$. The large field like contribution from BSOC would then lead to a large offset as shown in Fig. 4A only at t>$\lambda$sf with a field like torque approaching to zero at zero thickness. The contribution to measured field like torque reduces as R approaches 0. For a metal to metal surface R is usually a small number (*41*).

We use the following arguments against a BSOC only explanation of the measured data. First, $\tau_{fBSOC}$ will result in a thickness varying offset to the total field like torque. Second, the field like torque from BSOC ($\tau_{fBSOC}$) produces a measurable offset in the field like torque as R approaches one. Lastly, we show below that for a spherical Fermi surfaces R<0.5, we were unable to produce a large positive intercept ($H_{f\text{-intercept}}$ (t=0)) and a sign change for the total field like torque for R between 0.1 and 1. We further note that for a special case where all the Fermi surfaces are spherical and the same size where the spin dependent transmission is due to a surface potential (*40*),

$$\text{Re}(G) = \frac{1}{2} + \frac{u^\uparrow u^\downarrow}{2(u^\uparrow + u^\downarrow)}\left[u^\downarrow \ln\left(\frac{u^{\downarrow 2}}{1+u^{\downarrow 2}}\right) + u^\uparrow \ln\left(\frac{u^{\uparrow 2}}{1+u^{\uparrow 2}}\right)\right] \quad (12)$$

$$\text{Im}(G) = \frac{u^\uparrow u^\downarrow}{2(u^\uparrow + u^\downarrow)}\left(u^\downarrow\left[\pi - 2\tan^{-1} u^\downarrow\right] - u^\uparrow\left[\pi - 2\tan^{-1} u^\uparrow\right]\right) \quad (13)$$

where $u^\uparrow$, $u^\downarrow$ represent the strength of the spin dependent potential $V = (u_{\uparrow or \downarrow}\hbar^2 k_F/m)\delta(z)$ at the interface (*40,41*). The relation between the ratio $R = \text{Im}(G)/|G|$ and $u^\uparrow/u^\downarrow$ is plotted in Fig. 4E showing that under the spherical Fermi surface assumption R is <0.5. The typical experimental estimation of R for FM to NM interface is given in reference (*41*).

The experimentally measured field and damping torques can be explained using a semi-classical diffusive model that includes the presence of both BSOC and ISOC effects simultaneously. We also show three scenarios for the relative strength of the ISOC and BSOC effects in Fig. 5A, 5B, 5C. In all the scenarios, the measured field like torque exhibits a linear increase due to increasing current as the resistance of the HM electrode reduces with increasing thickness. The linear increase in the ampere field like torque with thickness is consistent with a constant resistivity as confirmed by 2-layer sheet resistance fits (Fig. S3). When BSOC is the only contribution to the spin torque (Fig. 5A), the measured field-like torque in mostly attributed to ampere torque. BSOC does generate a small field like torque (blue dotted line) due to the imaginary component of the spin mix conductance. We use an R=0.1 to include the effect of field like torque generated from BSOC. However, this scenario (BSOC only) does not produce an offset in the field like torque. When ISOC is the only contribution to the spin torque, the ISOC-damping torque is independent of thickness and has a small constant $-V_e$ value (*40*). Therefore, this scenario (ISOC only) does not explain the measured damping torque. When, both ISOC and BSOC are included in the model (Fig. 5B), the damping torque exhibits a saturating and increasing behavior consistent with measurement. The field like torque in the BSOC+ISOC scenario exhibits a linear increase in magnitude (due to an increasing Ampere field with increasing HM thickness) and a constant thickness independent off-set corresponding to an ISOC field-like contribution.

We fit the experimental data with the diffusive model to extract the relative strength of the ISOC and BSOC spin torques. The parameters used for the fitting are shown in Table 1. We estimate an electron mean free path of 0.5 nm based on the electron density of beta-tantalum ($5.58 \times 10^{22}$ cm$^{-3}$) assuming one valence electron per atom. Fig. 6A shows the fitting of the field like torque with an ampere torque combined with a large positive offset arising due to an interface spin orbit torque. Measurements were repeated at 7, 8. 9 and 10 GHz on all the seven samples. Clear sign change in measured Anti-symmetric component at 1.3 nm is observed for all 4 measurement frequencies.

The behavior of single layer CoFe at zero HM thickness exhibits a negative field like torque and a small positive damping torque. Non-zero Anti-symmetric component in the FMR signal for single layer FM has already been reported in a similar experiment and attributed to non-uniformity of FM (*42*) or a potential contribution from local spin orbit effects in FM (*41*). We also note that the sign of the anti-symmetric component observed at zero Ta thickness is opposite to the sign of the offset attributed to ISOC. We further note that Spin Pumping (SP) from the FM combined with Inverse Rashaba-Edelstein Effect (IREE)/Inverse Spin Hall Effect (ISHE) can produce a DC mixing voltage. However IREE & ISHE contribute only to the symmetric part of the FMR measurement (*43,44*) leading to a correction in the damping torques. Our conclusion of the presence of a strong field like ISOC contribution therefore should not be affected by the contribution from IREE/ISHE.

**References and Notes:**

(*1*) International Technology Roadmap for Semiconductors, 2013. [Online]. Available:

http://www.itrs.net/.

(*2*) K. Bernstein *et al.*, *Proc. IEEE* **98**, 2169–2184 (2010).

(*3*) D. Nikonov and I. Young, *Proceedings of IEDM*, 25.4 (2012).

(*4*) D. E. Nikonov and I. A. Young, *Proc. IEEE* **101**, 2498 - 2533 (2013).

(*5*) R. Sarpeshkar, *Ultra Low Power Bioelectronics*, Cambridge University Press (2010).

(*6*) H. Yoda *et al.*, *Proceedings of IEDM* 11.3 (2012).

(*7*) D. E. Nikonov and G. I. Bourianoff, *J. Supercond. Novel Magn.* **21**, 479–493 (2008).

(*8*) Y. Zhang *et al.*, *Design, Automation and Test in Europe Conference and Exhibition (DATE),2014,* 1-6, (2014).

(*9*) W. Kang, *et al.*, *SP-DAC* 676-683 (2014).

(*10*) W. J. Gallagher and S. S. P. Parkin, *IBM J. Res. and Dev.* **50**, 5-23 (2006).

(*11*) M. Tsoi et al., Phys. Rev. Lett. **80**, 4281 (1998).

(*12*) J. A. Katine et al., Phys. Rev. Lett. **84**, 3149 (2000).

(*13*) M. Hosomi, *et al.*, *IEDM Tech. Dig.* **459** (2005).

(*14*) M. I. Dyakonov and V. I. Perel, *Phys. Lett. A* **35**, 459 (1971).

(*15*) L. Liu *et al.*, *Phys. Rev. Lett.* **106**, 036601 (2011).

(*16*) C.-F. Pai *et al.*, *Appl. Phys. Lett.* **101**, 122404 (2012).

(*17*) L. Liu *et al.*, *Science* **336**, 555 (2012).

(*18*) A. R. Mellnik *et al.*, *Nature* **511**, 449 (2014).

(*19*) M. I. Dyakonov (ed.), *Spin physics in semiconductors*, **v. 157**, Berlin: Springer (2008).

(*20*) C. Kittel, *Phys. Rev* **71**, 270 (1947).

(*21*) T. R. McGuire and R. I. Potter, *IEEE Trans. Magn.* **11**, 1018 (1975).

(*22*) P. P. Freitas, L. Berger and J. F. Silvain, *J. Appl. Phys.* **61**, 4385 (1987).

(*23*) Y. Tserkovnyak, A. Brataas, and G. E. W. Bauer, *Phys. Rev. Lett.* **88**, 117601 (2002).

(*24*) H. Nakayama *et al.*, *Phys. Rev. Lett.* **110**, 206601 (2013).

(*25*) M. Morota *et al.*, *Phys. Rev. B* **83**, 174405 (2011).

(*26*) D. Bhowmik et al., *Proceedings of IEDM*, **29.7** (2012).

(*27*) C. Hahn *et al.*, *Phys. Rev. B* **87**, 174417 (2013).

(*28*) J. Kim *et al.*, *Nature Mat.* **12**, 240 (2013).

(*29*) H Nakayama *et al.*, *IEEE Trans. Magnet.* **46**, 2202 (2010).

(*30*) H. Nakayama *et al.*, *Phys. Rev. B* **85**, 144408 (2012).

(*31*) V. Castel *et al.*, *App. Phys. Lett.* **101**, 132414 (2012).

(*32*) J.-C. Rojas-Sánchez *et al.*, *Phys. Rev. Lett* **112**, 106602 (2014).

(*33*) A. Ganguly *et al.*, *App. Phys. Lett.* **104**, 072405 (2014).

(*34*) T. Yang *et al.*, *Jpn. J. App. Phys.* **53**, 04EM06 (2014).

(*35*) D. Hou *et al.*, *Appl. Phys. Lett.* **101**, 042403 (2012).

(*36*) A. Kovalev *et al.*, *Phys. Rev. Lett.* **75**, 014430 (2007).

(*37*) K. Oguz *et al.*, *J. Appl. Phys.* **103**, 07B526 2008

(*38*) K. Ando *et al.*, *Phys. Rev. Lett.* **101**, 036601 (2008).

(*39*) S. Petit *et al.*, *Phys. Rev. Lett.* **98**, 077203 (2007).

(*40*) P. M. Haney *et al.*, *Phys. Rev. B* **87**, 174411 (2013).

(*41*) A. Brataas, G. E. W. Bauer, and P. J. Kelly, *Phys. Rep.* **427**, 157–255 (2006).

(*42*) A. Tsukahara *et al.*, *Phys. Rev. B* **89**, 235317 (2014).

(*43*) Y. Tserkovnyak, A. Brataas, and G. E. W. Bauer. *Phys. Rev. B* **66** 224403 (2002).

(*44*) A. R, Mellnik *et al.*, *Nature* **511**, 449-451 (2014).

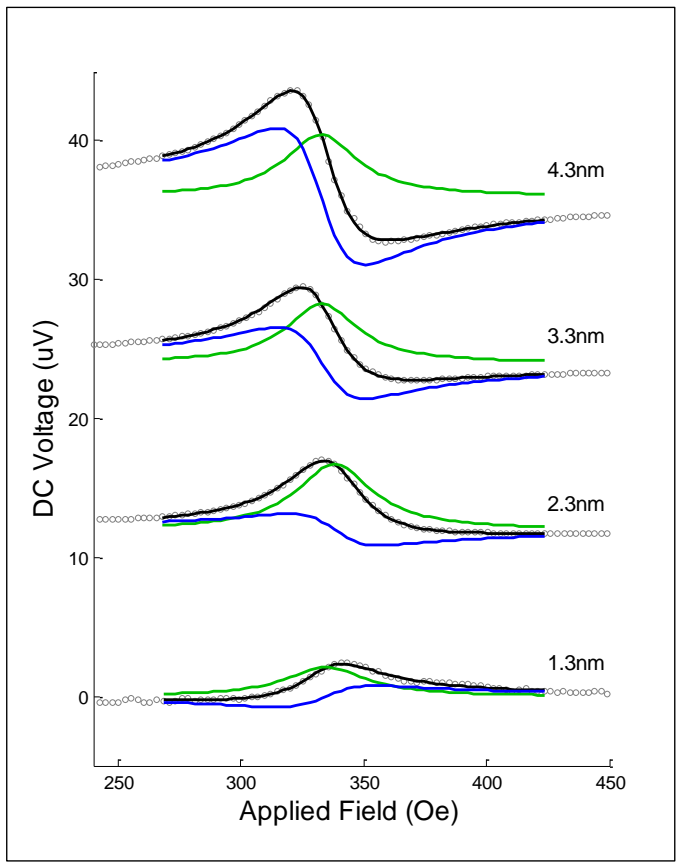

**Fig. 1.** DC voltage measurement for Ta on 4nm CoFeB. The labels refer to the Ta thickness. Shown are the measured data (black circles) and fit (black line) to data using equation (5), and the symmetric (green line) and anti-symmetric (blue line) components from the fit.

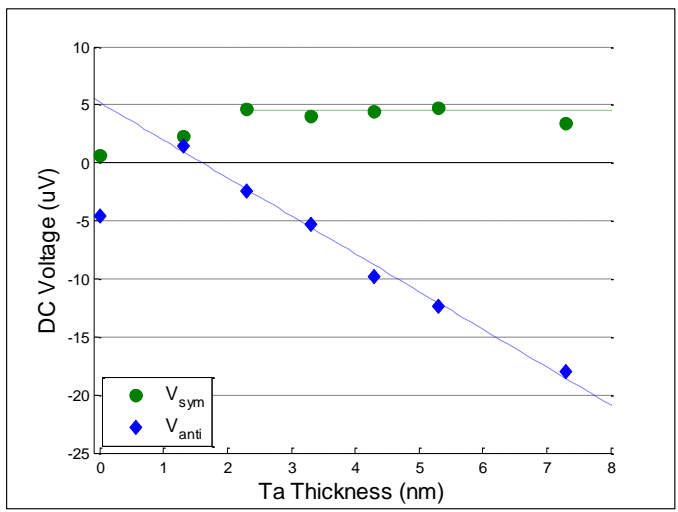

**Fig. 2.** Voltage of the symmetric (green circles) and anti-symmetric (blue diamonds) components versus Ta thickness on 4nm CoFeB at 7GHz. The data points at zero Ta thickness are for a single

6nm layer of CoFeB. The slope and intercept is a linear fit to $V_{anti}$ (dashed blue line) and the value of Vsym at the higher Ta thickness (dashed green line) are also shown.

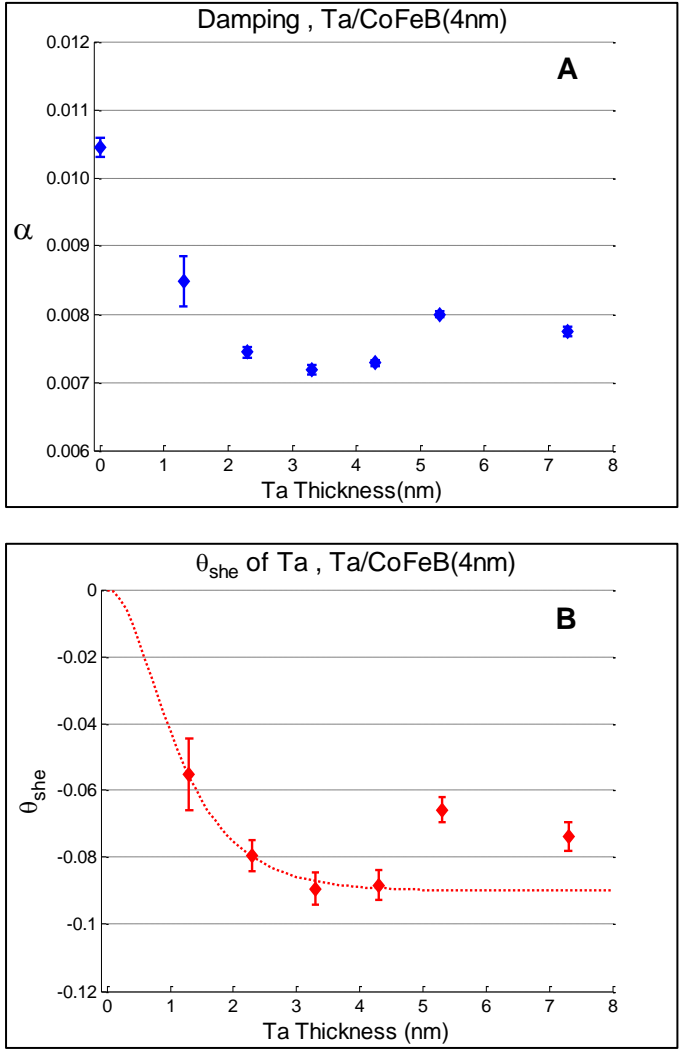

**Fig. 3.** (A) The measured damping coefficient, $\alpha$, versus Ta thickness. (B) The calculated value of SHE coefficient, $\theta_{sh}$, by assuming the decrease in $\alpha$ with Ta thickness is due to the spin torque from the SHE spin current of Ta.

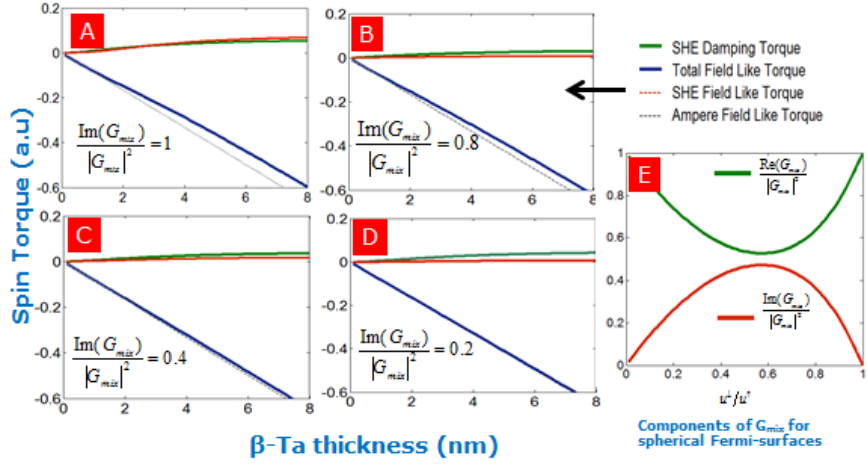

**Fig. 4.** BSOC (SHE) only scenario: Spin torque Vs thickness dependence if SHE and ampere field were the only spin/magnetic torques present in the system. (A) Ratio of the imaginary part of spin mix conductance R =1, (B) R=0.8, (C) R=0.4, (D) R=0.2. The field like torque produced by SHE (due to the presence of spin reflection at the HM/FM interface cannot produce a sign change in the field like torque combined with a positive constant offset. The field like torque from SHE approaches to zero for zero thickness. (E) Effect of varying interface potentials on the ratio of imaginary part of spin mixing conductance

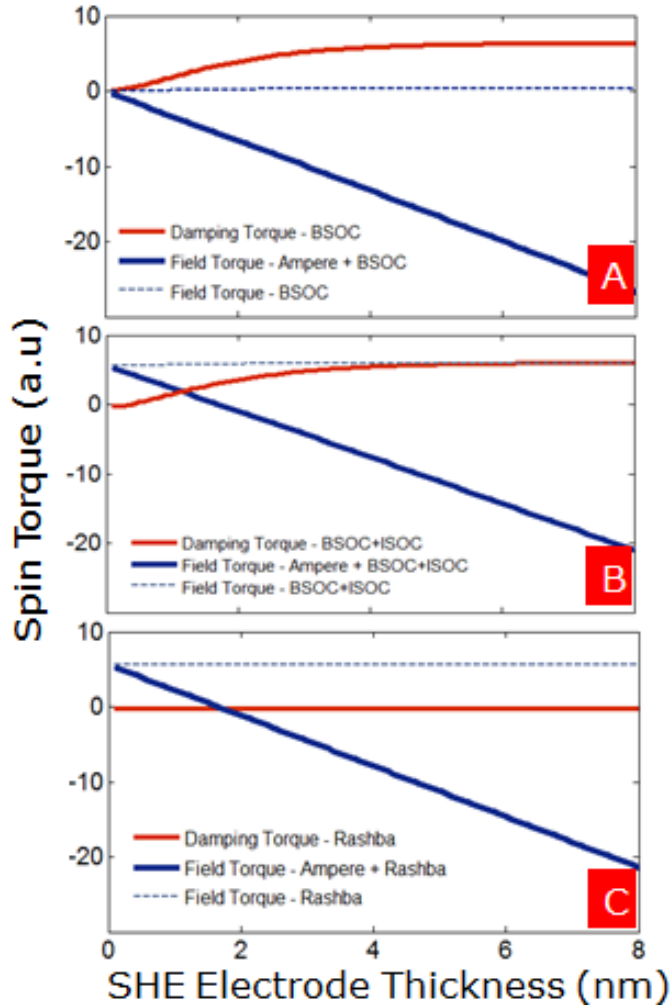

**Fig. 5.** Scenarios for varying relative strength of ISOC and BSOC. (A) BSOC only (B) ISOC and BSOC present in equal strengths (C) ISCO only.

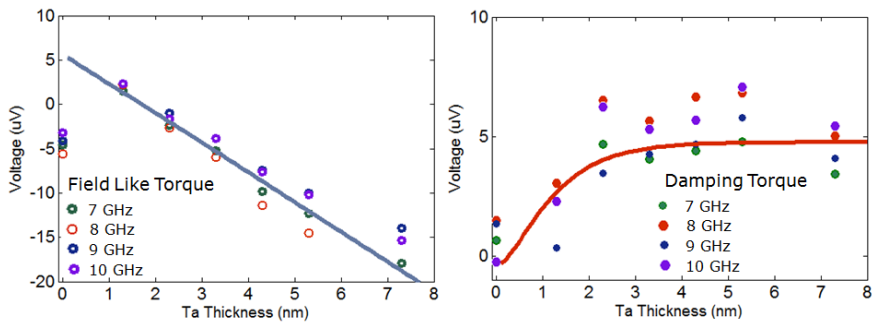

**Fig. 6.** ISOC+BOC based explanation for measured spin torques in CoFe/β-Ta bilayer. The measured torques are consistent with co-existence of strong field like interface spin orbit torque with a bulk spin orbit torque.

| Quantity | Value | Expression | Source/Ref. |
|---|---|---|---|
| Density (D) | 16.69 g/cm$^3$ | - | - |
| Resisitivity ($\rho_N$) | 185 $\mu\Omega\cdot$cm | Measured | - |
| Spin mixing conductance ($G_{mix}$) | 2.16x10$^{14}$ $\Omega^{-1}$/m$^2$ | Eq. 10 | (40) |
| Fermi wave-vector ($k_{F,\beta\text{-Ta}}$) | 11.8 nm$^{-1}$ | $k_f = (3\pi^2 D N_A N_V/Z)^{1/3}$ | (44) |
| Mean free path ($\lambda_{\beta\text{-Ta}}$) | 0.47 nm | $\lambda_n = (h/2e^2)\sigma^3 \pi k_F^{-2}$ | (44) |
| SHE coefficient | -0.11 | Eq. 7 | - |

**Table 1.** The parameters used for fitting to the experimental data to a diffusive model that includes the presence of both BSOC and ISOC.

# Supplementary Material for

# Experimental Demonstration of the Co-existence of the Spin Hall and Rashba Effects in beta-Tantalum/Ferromagnet Bilayers

**Authors:** Gary Allen[1*], Sasikanth Manipatruni[1], Dmitri E. Nikonov[1], Mark Doczy[1] and Ian A. Young[1]

**Affiliations**

[1] Components Research, Intel Corp., Hillsboro, Oregon, 97124, USA.

[*] Corresponding author. E-mail: gary.allen@intel.com

**Supplementary Materials:**

Materials and Methods

Figures S1-S4

**Materials and Methods**

The measured samples were composed of 20x100um Ta/CoFeB bilayer lines patterned as the center conductor of a coplanar waveguide with the Au ground lines and probe pads. The waveguide was patterned on 100nm of SiN on Silicon wafer substrate. The bilayers were grown by sputtering Ta and $Co_{20}Fe_{60}B_{20}$ (CoFeB) onto patterned resist followed by a resist lift off process. Ta was always deposited on top of CoFeB without an airbreak. The thicknesses of the Ta layer ranged between 1 to 8 nm, while the thickness of the CoFeB layer was kept constant at 4nm. Additionally, a sample with a single 6nm CoFeB layer was produced for comparison.

Thicknesses were determined by a combination of TEM images and resistivity measurements. Upon exposure to air, oxidation of the top Ta layer resulted in an oxide layer of approximately 3nm. Resistivities of 1850 Ω·nm and 1400 Ω·nm for Ta and CoFeB, respectively, were measured by sheet resistance measurements on blanket films. These values were used to calculate the resistances of the lines and matched well with DC resistance measurements when a contact resistance of approximately 50Ω is assumed (Fig. S3).

The large and constant resistivity found for all Ta thicknesses indicates that the films are entirely β-phase Ta, which is in line with our experience with Ta films less than 10nm in thickness. A magneto-resistance value of $2.1 \times 10^{-3} \pm 0.1 \times 10^{-3}$ was also determined from DC resistance measurements.

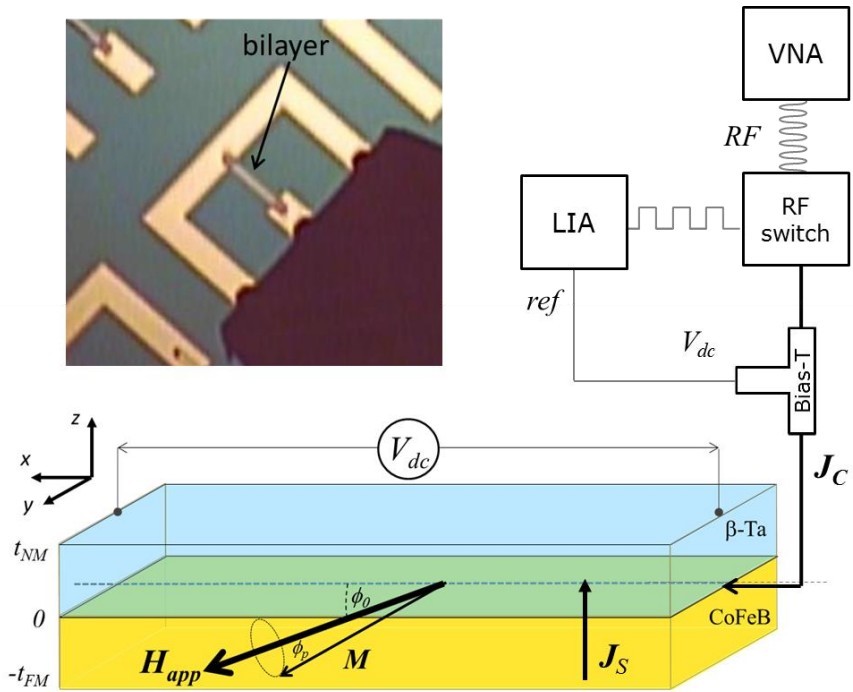

**Fig. S1.** Schematic of the FMR measurement. A vector network analyzer is used as an RF current source. The RF current is modulated by an RF switch and applied to a bilayer of β-Ta on CoFeB. An in-plane magnetic field is also applied to the bilayer at 45° to the direction of the charge current. A DC voltage results and is measured by a lock-in amplifier referenced to the frequency of the RF switch. In the photo is seen an RF probe in contact with the co-planar waveguide measurement structure with the Ta/CoFeB bilayer as the center conductor.

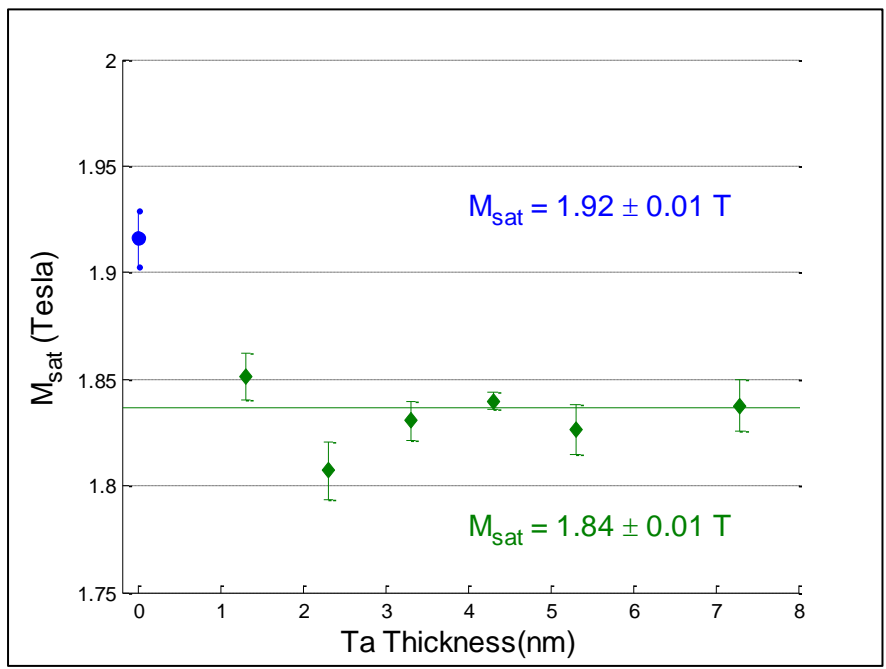

**Fig. S2.** The saturation magnetization, $M_{sat}$, determined for each sample.

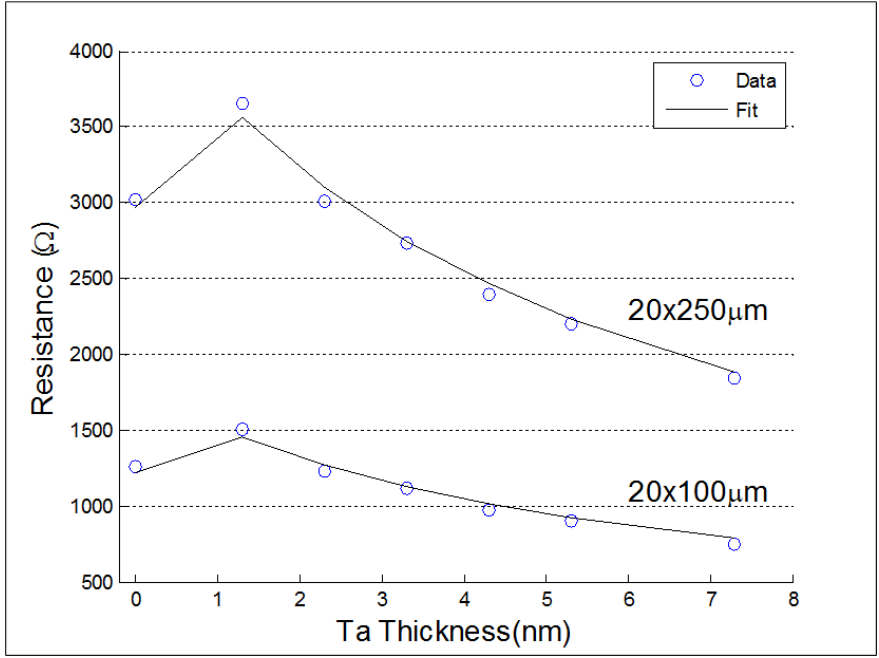

**Fig. S3.** The DC resistance of the Ta/CoFeB lines with 4nm of CoFeB. Data for both 100μm and 250μm length lines are shown. The circles represent measured data, while the solid lines are calculated from the film resistivities and thicknesses described in Materials and Methods. The data at zero Ta thickness is for a 6nm CoFeB layer without Ta.

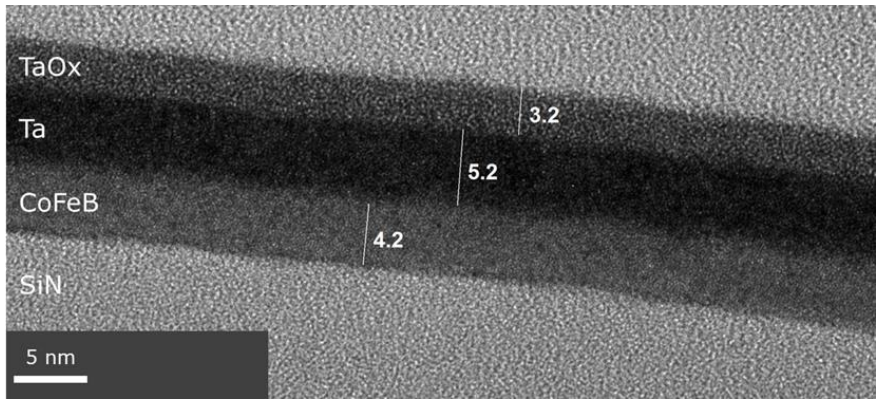

**Fig. S4.** TEM image of Ta/CoFeB bilayer on a Si nitride substrate. The Ta oxide layer is the native oxide of Ta that forms when exposed to air.